\documentclass[11pt,usletter]{article}
\usepackage{jcappub}
\pdfoutput=1

\usepackage{tensor}
\usepackage{graphicx}
\usepackage{color}
\usepackage{hyperref}

\begin{document}
\hfill{\small{FTUV-17-07-03, IFIC/17-37}}
\leftline{}

\title{Phantom Dirac-Born-Infeld Dark Energy}

\author{Gabriela Barenboim} 
\emailAdd{Gabriela.Barenboim@uv.es}
\affiliation{Departament de F\'isica Te\`orica and IFIC, Universitat de Val\`encia-CSIC, E-46100, Burjassot, Spain}

\author{William H.\ Kinney} 
\emailAdd{whkinney@buffalo.edu}
\author{Michael J. P. Morse} 
\emailAdd{mjmorse3@buffalo.edu}
\affiliation{Dept. of Physics, University at Buffalo,
        239 Fronczak Hall, Buffalo, NY 14260-1500}

\date{\today}

\abstract{
Motivated by the apparent discrepancy between Cosmic Microwave Background measurements of the Hubble constant and measurements from Type-Ia supernovae, we construct a model for Dark Energy with equation of state $w = p / \rho < -1$, violating the Null Energy Condition. Naive canonical models of so-called ``Phantom'' Dark Energy require a negative scalar kinetic term, resulting in a Hamiltonian unbounded from below and associated vacuum instability. We construct a scalar field model for Dark Energy with $w < -1$, which nonetheless has a Hamiltonian bounded from below in the comoving reference frame, {\it i.e.} in the rest frame of the fluid. We demonstrate that the solution is a cosmological attractor, and find that early-time cosmological boundary conditions consist of a ``frozen'' scalar field, which relaxes to the attractor solution once the Dark Energy component dominates the cosmological energy density. We consider the model in an arbitrary choice of gauge, and find that, unlike the case of comoving gauge, the fluid Hamiltonian is in fact unbounded from below in the reference frame of a highly boosted observer, corresponding to a nonlinear gradient instability. We discuss this in the context of general NEC-violating perfect fluids, for which this instability is a general property.
}

%\pacs{98.80.Cq}
\maketitle

\section{Introduction}
\label{sec:Intro}

Current cosmological data constraining the form of Dark Energy in the universe are consistent with Dark Energy as a cosmological constant \cite{Ade:2015xua}. However, recent direct measurements of the Hubble parameter $H_0$ are in substantial tension with measurements based on the Cosmic Microwave Background (CMB) \cite{Ade:2015xua,Riess:2016jrr,Bonvin:2016crt,Bernal:2016gxb,DiValentino:2016hlg,Lukovic:2016ldd,Archidiacono:2016kkh,Tram:2016rcw,Aghanim:2016sns,Ko:2016uft,Zhao:2016ecj,Cardona:2016ems,Lin:2017ikq,Freedman:2017yms,Abbott:2017wau,Joudaki:2017zdt,Zhang:2017aqn,Feeney:2017sgx,Addison:2017fdm}. While perhaps the most parsimonious explanation of this tension is the presence of an unidentified systematic in one or more data sets \cite{Bernal:2016gxb,Romano:2016utn,Shafieloo:2016zga,Calcino:2016jpu,Lattanzi:2016dzq,Fleury:2016fda,Odderskov:2017ivg,Heavens:2017hkr,Wu:2017fpr,Jesus:2017zej,Lin:2017bhs,Hou:2017smn,Aylor:2017haa,Follin:2017ljs}, the possibility remains that the tension in $H_0$ between high-redshift and low-redshift measurements is an indication of new physics beyond the six-parameter ``concordance'' model of cosmology. Possibilities for this new physics include ``dark radiation'', {\it i.e.} an extra light degree of freedom \cite{Riess:2016jrr,Sasankan:2016ixg,Feng:2017mfs,DiValentino:2016ucb,Barenboim:2016lxv,Gerbino:2016sgw,Verde:2016wmz,Benetti:2017gvm,Feng:2017nss,Zhao:2017urm,Brust:2017nmv,Gariazzo:2017pzb,Sasankan:2017eqr}, dynamical or interacting Dark Energy \cite{Riess:2016jrr,Farooq:2016zwm,Xu:2016grp,Xia:2016vnp,Kumar:2016zpg,Karwal:2016vyq,Chacko:2016kgg,Marcondes:2016zte,vandeBruck:2016hpz,Murgia:2016dug,Sola:2016hnq,Zhao:2017cud,Kumar:2017dnp,Sola:2017jbl,Zhang:2017idq,DiValentino:2017zyq,Camera:2017tws,DiValentino:2017iww,CarrilloGonzalez:2017cll,Dhawan:2017leu,Sola:2017znb,Yang:2017yme,Magana:2017nfs,Lonappan:2017lzt,vanPutten:2017qte,Miranda:2017rdk}, and nonzero curvature \cite{Farooq:2016zwm,Bolejko:2017gnj,Ooba:2017ukj,Bolejko:2017wfy}. In this paper, we concentrate on the possibility of a ``phantom'' equation of state for Dark Energy \cite{Caldwell:1999ew,Caldwell:2003vq,Dabrowski:2004hx,Alam:2016wpf,Riess:2016jrr,DiValentino:2016hlg}, which corresponds to Dark Energy equation of state $w \equiv p / \rho < -1$, violating the Null Energy Condition (NEC). Phantom Dark Energy provides an especially simple resolution to the discrepancy in measurements of $H_0$: high-reshift measurements favor a small value of $H_0$, and low-redshift measurements favor a larger value, which can be readily explained by an increasing expansion rate, corresponding to equation of state $w < -1$.  Constraints on Phantom Dark Energy (PDE) were calculated, {\it e.g.}, by Di Valentino and Silk in Ref. \cite{DiValentino:2016hlg}, with a 68\% confidence level constraint of $w = -1.29^{+0.15}_{-0.12}$, using the Planck CMB measurement \cite{Ade:2015xua} and the Riess {\it et al.} constraint on $H_0$ from Type-Ia supernova data \cite{Riess:2016jrr}. (While inclusion of PDE improves the fit relative to the $\Lambda$CDM case, we note that the extended parameters are nonetheless disfavored by Bayesian evidence \cite{Zhao:2017cud,Ooba:2018dzf}. In this paper, we adopt the best-fit from Ref. \cite{DiValentino:2016hlg} as a fiducial case {\it consistent} with current data, although not yet convincingly favored over $\Lambda$CDM.)

While appealing from a parametric standpoint, Phantom Dark Energy is less so from the standpoint of fundamental physics, since NEC violation in scalar field theory requires a negative kinetic term in the field Lagrangian, for example in the simplest canonical realization \cite{Faraoni:2005gg},
\begin{equation}
\label{eq:canonicalphantom}
{\mathcal L} = - X - V\left(\phi\right),
\end{equation} 
where
\begin{equation}
X \equiv \frac{1}{2} g_{\mu \nu} \partial^\mu \phi \partial_\nu \phi.
\end{equation}
Therefore the corresponding Hamiltonian, corresponding to the field energy density, is unbounded from below,
\begin{equation}
{\mathcal H} = - X + V\left(\phi\right),
\end{equation}
so that ${\mathcal H} \rightarrow - \infty$ as $X \rightarrow \infty$. The result is vacuum instability via particle creation, and a theory which is not self-consistent \cite{Carroll:2003st,Cline:2003gs,Dubovsky:2005xd}. In addition, Phantom Dark Energy results in a future cosmological singularity, or `Big Rip' \cite{Caldwell:2003vq}. The literature on proposed solutions to these problems is large. In this paper, we consider an especially simple  approach to the instability problem by considering a phenomenological Lagrangian of the Dirac-Born-Infeld (DBI) form, which reduces to the form of a canonical phantom field (\ref{eq:canonicalphantom}) in the $X \rightarrow 0$ limit, but nonetheless has a postive-definite comoving energy density in the $X \rightarrow \infty$ limit. The toy model we consider has constant equation of state, $w = {\mathrm const.} < 0$, which can serve to alleviate the tension in $H_0$ between low-redshift and high-redshift constraints, but still has the issue of a future Big Rip singularity. While the energy density of the field is in bounded below in the comoving reference frame, we show that this property does not apply to the Hamiltonian evaluated in arbitrary gauge, and that it is always possible to construct a gauge in which the Hamiltonian is in fact unbounded from below, indicating an instability in the theory. This instability is in fact characteristic of general NEC-violating perfect fluids, a result which was first shown by Sawiki and Vikman \cite{Sawicki:2012pz} which we summarize in Sec. \ref{sec:ArbitraryGauge}.

The paper is organized as follows: In Section \ref{sec:GeneralFormalism}, we review the non-canonical ``flow'' formalism which we use to construct the Phantom Dark Energy solution. In Section \ref{sec:DarkEnergyModel} we construct a stable Phantom Dark Energy Lagrangian. In Section \ref{sec:FieldDynamics} we demonstrate that the solution is a general dynamical attractor, and consider early-universe boundary conditions and vacuum stability in arbitrary gauge. Section \ref{sec:Conclusions} presents conclusions. 

\section{General Formalism}
\label{sec:GeneralFormalism}

In this section, we briefly review the ``flow'' formalism for non-canonical Lagrangians, following closely the discussion in Bean {\it{et al.}} \cite{Bean:2008ga} and Bessada, {\it{et al.}} \cite{Bessada:2009ns}. We will use this formalism in Sec. \ref{sec:DarkEnergyModel} to construct an exactly solvable scalar Dark Energy model with $w < -1$. 

We take a general Lagrangian of the form
${\cal{L}}={\cal{L}}\left[X,\phi\right]$, where $2X=g^{\mu\nu}\partial_{\mu}\phi\partial_{\nu}\phi$ is the canonical kinetic term. We assume a flat Friedmann-Robertson-Walker metric of the form
\begin{equation}
g_{\mu\nu} = \mathrm{diag.}(1, -a^2(t), -a^2(t), -a^2(t)),
\end{equation}
so that $X$ is positive-definite. The pressure $p$ and energy density $\rho$ are given by
\begin{eqnarray}
p &=& {\cal L}\left(X,\phi\right), \label{eq:p}\\
\rho &=& 2 X {\cal L}_X - {\cal L}, \label{eq:rho}
\end{eqnarray}
where the subscript ``${X}$" indicates a derivative with respect to the kinetic term.  The Friedmann equation can be written in terms of the reduced Planck mass  $M_P \equiv 1/\sqrt{8\pi G}$,
\begin{equation}
H^2 = \left(\frac{\dot a}{a}\right)^2 = \frac{1}{3 M_P^2} \rho = \frac{1}{3 M_P^2} \left(2X {\cal
L}_X - {\cal L}\right),
\end{equation}
and stress-energy conservation results in the continuity equation,
\begin{equation}
\dot\rho = 2 H {\dot H} = -3 H \left(\rho + p\right) = - 6 H X {\cal L}_X.
\end{equation}
For monotonic field evolution, the field value $\phi$ can be used as a ``clock'', and all other quantities expressed as functions of $\phi$, for
example $X = X\left(\phi\right)$, ${\cal L} = {\cal L}\left[X\left(\phi\right),\phi\right]$, and so on. We consider the homogeneous case, so that
$\dot\phi = \sqrt{2 X}$. Using
\begin{equation}
\frac{d}{dt} = \dot\phi \frac{d}{d \phi} = \sqrt{2 X} \frac{d}{d\phi},
\end{equation}
we can re-write the Friedmann and continuity equations as the {\it Hamilton Jacobi} equations,
\begin{eqnarray}
\label{eq:hamjac1} \dot
\phi = \sqrt{2 X} &=& -\frac{2M_P^2}{{\cal L}_{X}}H'(\phi),\\ \label{eq:hamjac2}
3M_P^2H^2(\phi)&=&\frac{4M_P^4{H'\left(\phi\right)}^2}{{\cal L}_{X}}-{\cal L}.
\end{eqnarray}
where a prime denotes a derivative with respect to the field $\phi$.

We define {\it flow parameters} as derivatives with respect to the number of e-folds, $dN \equiv d \log{a(t)} = H dt$:
\begin{eqnarray}
\label{eq:epsN}
\epsilon &\equiv& - \frac{1}{H} \frac{d H}{dN}
,\\ \label{eq:sN}
s &\equiv& \frac{1}{c_S} \frac{d c_S}{dN},\\ \label{eq:stilN}
\tilde{s} &\equiv& - \frac{1}{{\cal{L}}_{X}}\frac{d {\cal{L}}_{X}}{dN}.
\end{eqnarray}
where speed of sound for the scalar fluid is given by
\begin{equation}
\label{eq:defspeedofsound} c_S^{2} \equiv \frac{p_X}{\rho_X} = \left(1 + 2X\frac{{\cal
L}_{XX}}{{\cal L}_{X}}\right)^{-1},
\end{equation}
(Note that we adopt the opposite sign convention for $N$ than used {\it e.g.} in Ref. \cite{Bessada:2009ns}, appropriate to late-time cosmic acceleration.) 
The equation of state of the scalar field $\phi$ is related to the parameter $\epsilon$ by:
\begin{equation}
\label{eq:defw}
w \equiv \frac{p}{\rho} = \frac{2}{3} \epsilon - 1. 
\end{equation}

For monotonic field evolution, number of e-folds $dN$ can then be re-written in terms of $d\phi$ by:
\begin{eqnarray}
\label{eq:Nphi}
dN \equiv H dt &&= \frac{H}{\sqrt{2 X}} d\phi\\
&&= - \frac{{\cal L}_X}{2 M_P^2} \left(\frac{H\left(\phi\right)}{H'\left(\phi\right)}\right) d\phi,
\end{eqnarray}
and the flow parameters $\epsilon$, $s$, and $\tilde s$ (\ref{eq:epsN}) can be written as derivatives with respect to the field $\phi$ as \cite{Bean:2008ga}:
\begin{eqnarray}
\label{eq:defeps}
\epsilon\left(\phi\right) &=& \frac{2
M_P^2}{{\cal{L}}_{X}}
\left(\frac{H'\left(\phi\right)}{H\left(\phi\right)}\right)^2,\\ \label{eq:defs}
s\left(\phi\right) &=& - \frac{2
M_P^2}{{\cal{L}}_{X}}
\frac{H'\left(\phi\right)}{H\left(\phi\right)} \frac{c_S'\left(\phi\right)}{c_S\left(\phi\right)},\\ \label{eq:defstil}
\tilde{s}\left(\phi\right) &=& \frac{2
M_P^2}{{\cal{L}}_{X}}
\frac{H'\left(\phi\right)}{H\left(\phi\right)}\frac{{\cal{L'}}_{X}}{{\cal{L}}_{X}}.
\end{eqnarray}
Following Refs. \cite{Kinney:2007ag,Bessada:2009ns}, we construct a family of exact solutions by making the {\it ansatz} of $\epsilon$, $s$, and $\tilde s$ constant, so that 
\begin{eqnarray}
\label{eq:solN}
H &\propto& e^{-\epsilon N}\cr
c_S &\propto& e^{s N},\cr
{\mathcal L}_X &\propto& e^{-{\tilde s} N}.
\end{eqnarray}
We can write these expressions as solutions to Eqs. (\ref{eq:Nphi}, \ref{eq:defeps}, \ref{eq:defs}, \ref{eq:defstil}) as follows: 
\begin{eqnarray}
\label{eq:solphi}
&&\phi\left(N\right) = \phi_0 e^{{\tilde s} N / 2},\cr
&&c_S\left(\phi\right) = \left(\frac{\phi}{\phi_0}\right)^{2 s / {\tilde s}},\cr
&&H\left(\phi\right) = H_0 \left(\frac{\phi}{\phi_0}\right)^{-2 \epsilon / s},\cr
&&{\mathcal L}_X = \frac{8 \epsilon}{{\tilde s}^2} \left(\frac{M_{\rm P}}{\phi}\right)^2.
\end{eqnarray}
Here the field value $\phi_0$ is defined such that $c_S\left(\phi_0\right) = 1$, and the solution admits both causal ($c_S < 1$) and ``tachyacoustic'' ($c_S > 1$) behavior. Note in particular that we have not yet specified the form of the Lagrangian leading to solutions of the form (\ref{eq:solphi}): In fact, such solutions define a family of Lagrangians, which are determined via the relationship between the parameters $s$ and ${\tilde s}$. (See Ref. \cite{Bessada:2009ns} for a detailed discussion.) For our purposes here, it is sufficient to specify a Lagrangian ${\mathcal L}$ by {\it ansatz}, and demonstrate that it admits a solution of the form (\ref{eq:solphi}). In the next section, we construct a DBI-like Lagrangian with solution (\ref{eq:solphi}) characterized $\epsilon < 0$, corresponding to $w < -1$. 

\section{Dark Energy Model}
\label{sec:DarkEnergyModel}

In this section, we construct a general DBI-like model with constant equation of state $w < -1$. Consider a Lagrangian of the form
\begin{equation}
\label{eq:DBILag}
{\mathcal L} = -\frac{1}{f\left(\phi\right)} \sqrt{1 \pm 2 f\left(\phi\right) X} +\frac{1}{f\left(\phi\right)} - V\left(\phi\right).
\end{equation}
It is conventional to define the Lagrangian such that the limit $X \rightarrow 0$ corresponds to a canonical Lagrangian, 
\begin{equation}
{\mathcal L} \rightarrow  X - V\left(\phi\right),
\end{equation}
which is equivalent to choosing the negative sign in Eq. (\ref{eq:DBILag}). This is the standard Dirac-Born-Infeld  case. 
Here we make the opposite {\it ansatz}, 
\begin{equation}
\label{eq:PhanLag}
{\mathcal L} = -\frac{1}{f\left(\phi\right)} \sqrt{1 + 2 f\left(\phi\right) X} + \frac{1}{f\left(\phi\right)} - V\left(\phi\right),
\end{equation}
with $f\left(\phi\right) > 0$, so that the ``canonical'' limit has a wrong-sign kinetic term as $X \rightarrow 0$, 
\begin{equation}
{\mathcal L} \rightarrow  -X - V\left(\phi\right).
\end{equation}
Despite the wrong-sign kinetic term, the energy density (\ref{eq:rho}) corresponding to the Lagrangian (\ref{eq:PhanLag}) is bounded from below for an appropriate choice of potential $V$:
\begin{equation}
\label{eq:rho2}
\rho = \frac{1}{f\left(\phi\right) \sqrt{1 + 2 f\left(\phi\right) X}} - \frac{1}{f\left(\phi\right)} + V\left(\phi\right),
\end{equation}
which is positive definite as long as $V\left(\phi\right) > f\left(\phi\right)^{-1}$ for all values of the field $\phi$. 
The speed of sound (\ref{eq:defspeedofsound}) is
\begin{equation}
\label{eq:DBIcS}
c_S = + \sqrt{1 + 2 f\left(\phi\right) X} = - \frac{1}{{\mathcal L}_X}. 
\end{equation}
Comparing with Eqs. (\ref{eq:defs}) and (\ref{eq:defstil}), we then have immediately that $\tilde s = s$. The Hamilton-Jacobi Equations (\ref{eq:hamjac1},\ref{eq:hamjac2}) reduce to:
\begin{eqnarray}
\label{eq:DBIHJ1}
&&\dot\phi = \sqrt{2 X} = 2 M_{\rm P}^2 c_S\left(\phi\right) H'\left(\phi\right),\\ \label{eq:DBIHJ2}\cr
&& {\mathcal L} = - 3 M_{\rm P}^2 H^2 \left(1 - \frac{2 \epsilon}{3}\right).
\end{eqnarray}
Note in particular that the field evolution is in the direction of {\it increasing} Hubble parameter, $\dot\phi \propto + H'\left(\phi\right)$.

We wish to construct functions $f\left(\phi\right)$ and $V\left(\phi\right)$ which admit solutions of the form (\ref{eq:solphi}), with $\epsilon$ and $s$ constant, and $w < -1$, so that $\epsilon < 0$. From Eqs. (\ref{eq:DBIcS}), (\ref{eq:DBIHJ1}), and (\ref{eq:defeps}), we can construct the functional form of $f\left(\phi\right)$,
\begin{equation}
\label{eq:DBIf}
f = \frac{1}{2 M_{\rm P}^2 H^2 \epsilon} \frac{1 - c_S^2\left(\phi\right)}{c_S\left(\phi\right)}. 
\end{equation}
Writing the Lagrangian as
\begin{equation}
{\mathcal L} = \frac{1}{f} \left(1 - c_S\right) - V\left(\phi\right),
\end{equation}
the Hamilton-Jacobi Equation (\ref{eq:DBIHJ2}), combined with the solution (\ref{eq:DBIf}) for $f$ results in an expression for $V\left(\phi\right)$,
\begin{equation}
V\left(\phi\right) = 3 M_{\rm P}^2 H^2\left(\phi\right) \left[1 -\frac{2 \epsilon}{3} \left(\frac{1}{1 + c_S\left(\phi\right)}\right)\right].
\end{equation}
We make contact with the ansatz (\ref{eq:solphi}) by taking
\begin{eqnarray}
\label{eq:gensol1}
&&\phi\left(N\right) = \phi_0 e^{s N / 2},\cr
&&c_S\left(\phi\right) = \left(\frac{\phi}{\phi_0}\right)^2,\cr
&&H\left(\phi\right) = H_0 \left(\frac{\phi}{\phi_0}\right)^{-2 \epsilon / s},
\end{eqnarray}
so that $f\left(\phi\right)$ and $V\left(\phi\right)$ take the functional forms,
\begin{equation}
\label{eq:gensol2}
f\left(\phi\right) = \frac{1}{2 M_{\rm P}^2 H_0^2 \epsilon} \left(\frac{\phi}{\phi_0}\right)^{4 \epsilon / s - 2} \left[1 - \left(\frac{\phi}{\phi_0}\right)^4\right],
\end{equation}
and
\begin{equation}
\label{eq:gensol3}
V\left(\phi\right) = 3 M_{\rm P}^2 H_0^2 \left(\frac{\phi}{\phi_0}\right)^{-4 \epsilon / s} \left[1 - \frac{2 \epsilon}{3} \frac{1}{1 - \left(\phi / \phi_0\right)^2}\right].
\end{equation}
It is straightforward to verify that Eqs. (\ref{eq:gensol1}), (\ref{eq:gensol2}), (\ref{eq:gensol3}) satisfy the Hamilton-Jacobi Equations (\ref{eq:DBIHJ1}) and (\ref{eq:DBIHJ2}).

\begin{figure}
\begin{center}
\includegraphics[width=0.8\textwidth]{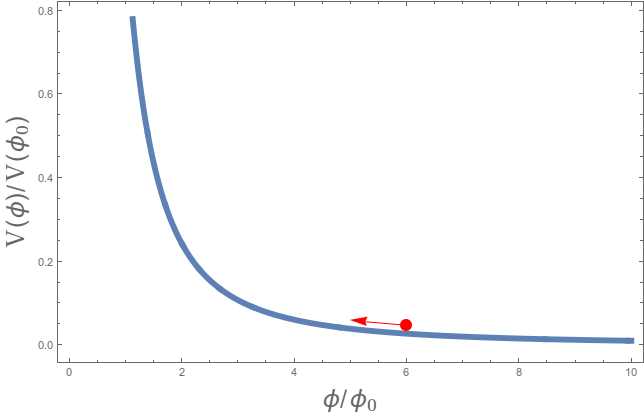}
\end{center}
\caption{The potential for $s = 2 \epsilon$, showing field evolution.}
\label{fig:potential}
\end{figure}

The solution (\ref{eq:gensol1},\ref{eq:gensol2},\ref{eq:gensol3}) represents a family of Dark Energy models parameterized by $\epsilon$ and $s$. The parameter $\epsilon$ is directly related to the equation of state parameter $w$ by Eq. (\ref{eq:defw}), but the parameter $s$ is arbitrary, and can be chosen to obtain a conceptually simple Dark Energy model. (We consider one such example here, although others are possible.) Take the case of 
\begin{equation}
s = 2 \epsilon,
\end{equation}
so that
\begin{eqnarray}
\label{eq:s2e}
&&f\left(\phi\right) =  \frac{1}{2 M_{\rm P}^2 H_0^2 \epsilon} \left[1 - \left(\frac{\phi}{\phi_0}\right)^4\right],\cr
&&V\left(\phi\right) = 3 M_{\rm P}^2 H_0^2 \left(\frac{\phi}{\phi_0}\right)^{-2}  \left[1 - \frac{2 \epsilon}{3} \frac{1}{1 + \left(\phi / \phi_0\right)^2}\right],
\end{eqnarray}
with solution
\begin{eqnarray}
\label{eq:s2esol1}
H\left(\phi\right) &&= H_0 \left(\frac{\phi}{\phi_0}\right)^{-1},\cr
c_S &&= \left(\frac{\phi}{\phi_0}\right)^2,
\end{eqnarray}
and field velocity
\begin{eqnarray}
\label{eq:s2esol2}
\dot\phi &&= 2 M_{\rm P}^2 c_S H'\left(\phi\right)\cr
         &&= - \frac{2 M_{\rm P}^2 H_0}{\phi_0} = {\rm const.}
\end{eqnarray}
We then have an approximately quadratically declining potential, $V \propto \phi^{-2}$, with field rolling {\it up} the potential \cite{Csaki:2005vq,Sahlen:2005zw} with constant velocity $\dot\phi = {\rm const.} < 0$ (Fig. \ref{fig:potential}). We will use this as a Dark Energy model.\footnote{Note that this model is purely phenomenological: a wrong-sign DBI Lagrangian of the type we propose is unlikely to arise in realistic string or braneworld models for ultraviolet (UV) physics. The question of a self-consistent UV completion resulting in a low-energy effective Lagrangian of the form (\ref{eq:PhanLag}) is an interesting one, but is beyond the scope of this work.}

%%%%%%%%%%%%%%%%%%%%%%%%%%%%%%%%%%%%%%%%%%%%%%%%%%%%%%%%%%%%%%%%%%%%%%%%%%%%%%%%%%%%%%%%%%%%%%%%%
%%%%%%%%%%%%%%%%%%%%%%%%%%%%%%%%%%%%%%%%%%%%%%%%%%%%%%%%%%%%%%%%%%%%%%%%%%%%%%%%%%%%%%%%%%%%%%%%%

In the next section, we demonstrate that this solution is a dynamical attractor, and discuss cosmological boundary conditions.

\section{General Field Dynamics}
\label{sec:FieldDynamics}

\subsection{Attractor Behavior}
\label{sec:AttractorBehavior}

While it is straightforward to demonstrate that Eqs. (\ref{eq:gensol1},\ref{eq:gensol2},\ref{eq:gensol3}) represent {\it a} solution for field evolution in a Lagrangian of the form (\ref{eq:PhanLag}), it is not immediately clear that this solution represents a dynamical attractor, which is necessary for the construction of a viable Dark Energy model. In this section, we demonstrate that the solution is, in fact, a dynamical attractor. 

The equation of motion for a Lagrangian of the form (\ref{eq:PhanLag}) can be shown to be:
\begin{equation}
\ddot\phi + 3 \frac{H}{\gamma^2} \dot\phi + \frac{3}{2} \frac{f'\left(\phi\right)}{f\left(\phi\right)} \dot\phi^2 +  \frac{f'\left(\phi\right)}{f\left(\phi\right)^2} -  \frac{1}{\gamma^3} \left[\frac{f'\left(\phi\right)}{f\left(\phi\right)^2} + V'\left(\phi\right)\right] = 0,
\end{equation}
where 
\begin{equation}
\gamma \equiv c_s^{-1} = \frac{1}{\sqrt{1 + f\left(\phi\right) \dot\phi^2}}.
\end{equation}
We can write this in dimensionless phase-space variables as follows:
Take 
\begin{equation}
x \equiv \frac{\phi}{\phi_0},
\end{equation}
and
\begin{equation}
\label{eq:psmom}
y = y(x) = \frac{\dot\phi}{\sqrt{3} M_{\rm P} H_0},
\end{equation}
where we take $x(t)$ to be monotonic, so $y(t) = y[x(t)] = y(x)$. We can likewise define dimensionless forms for the warp factor (\ref{eq:gensol2}) and potential (\ref{eq:gensol3}) as
\begin{equation}
\label{eq:dimf}
g\left(x\right) \equiv 3 M_{\rm P}^2 H_0^2 f\left(\phi\right) = \frac{3}{2 \epsilon} x^{4 \epsilon / s - 2} \left(1 - x^4\right),
\end{equation}
and 
\begin{equation}
\label{eq:dimv}
v(x) \equiv \frac{V\left(\phi\right)}{3 M_{\rm P}^2 H_0^2} = x^{-4 \epsilon/s} \left(1 - \frac{2 \epsilon}{3} \frac{1}{1 + x^2}\right).
\end{equation}
Using (\ref{eq:psmom}), we can write
\begin{equation}
\ddot\phi = \dot\phi \frac{d \dot\phi}{d \phi} = \frac{3 M_{\rm P}^2 H_0^2}{\phi_0} y(x) y'(x),
\end{equation}
and defining
%
%\begin{equation}
%h(x)^2 \equiv 3 M_{\rm P}^2 H^2,
%\end{equation}
%
\begin{equation}
h(x) \equiv H / H_0,
\end{equation}
we can write a general dimensionless equation for evolution of the system in phase space, appropriate for numerical evolution:
\begin{equation}
\label{eq:dimeom}
y(x) y'(x) + 3 \sqrt{- \frac{8 \epsilon}{3 s^2}} \frac{h(x) y(x)}{\gamma^2\left(x\right)} + \frac{g'\left(x\right)}{g\left(x\right)} \left[\frac{3}{2} g\left(x\right) y^2\left(x\right) + \left(1 - \frac{1}{\gamma^3\left(x\right)}\right)\right] - \frac{v'\left(x\right)}{\gamma^3\left(x\right)} = 0.
\end{equation}
The analytic solution (\ref{eq:gensol1}) then corresponds to
\begin{eqnarray}
c_S\left(x\right) &&= \gamma^{-1}\left(x\right) = x^2,\cr
h\left(x\right) &&= x^{-2 \epsilon / s},\cr
y\left(x\right) && = \pm   \sqrt{- \frac{2 \epsilon}{3}} x^{1 - 2 \epsilon / s},
\end{eqnarray}
where the sign of $y\left(x\right)$ is the same as the sign of $s$. It is straightforward to verify that this is an exact solution to Eq. (\ref{eq:dimeom}). 

\begin{figure}
\begin{center}
\includegraphics[width=0.8\textwidth]{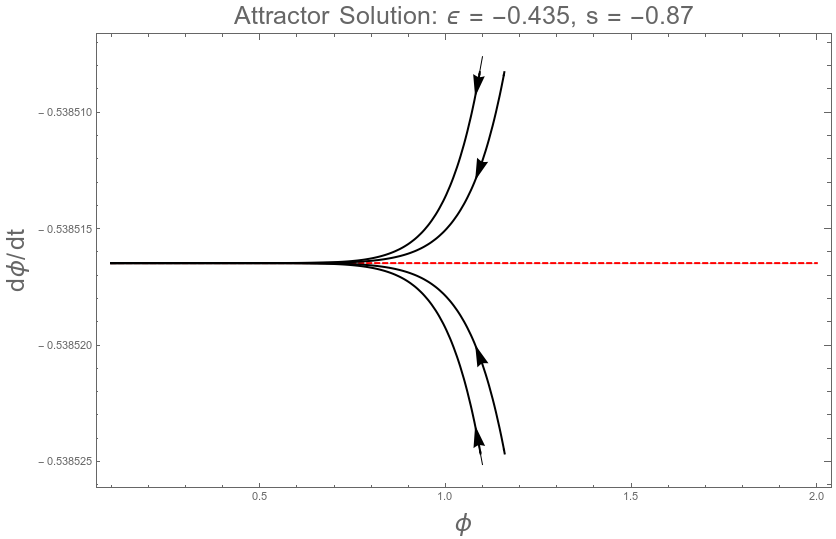}
\end{center}
\caption{Phase space plot of attractor behavior for the phantom scalar field. The dotted (red) line represents the late-time attractor solution, and the solid lines represent exact numerical solutions for the field evolution. Note that the field is rolling {\it up} the potential, with $\dot\phi < 0$.}
\label{fig:attractor1}
\end{figure}

\begin{figure}
\begin{center}
\includegraphics[width=0.8\textwidth]{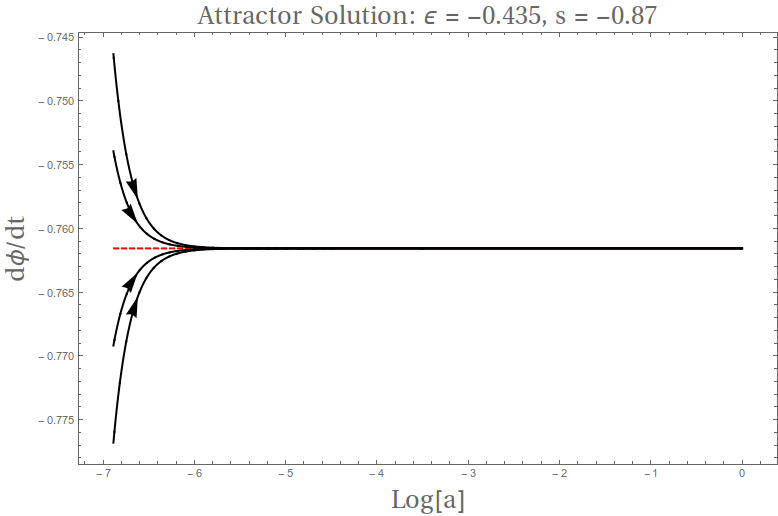}
\end{center}
\caption{Attractor behavior for the phantom scalar field, plotted vs. scale factor $a$. The dotted (red) line represents the late-time attractor solution, and the solid lines represent exact numerical solutions for the field evolution.}
\label{fig:attractor2}
\end{figure}

We are particularly interested in the case $s = 2 \epsilon$, (\ref{eq:s2e},\ref{eq:s2esol1},\ref{eq:s2esol2}), which corresponds to
\begin{eqnarray}
&&g\left(x\right) = \frac{3}{2 \epsilon} \left(1 - x^4\right),\cr
&&v(x) = x^{-2} \left(1 - \frac{2 \epsilon}{3} \frac{1}{1 + x^2}\right),
\end{eqnarray}
with analytic solution
\begin{eqnarray}
\label{eq:attractor}
c_S\left(x\right) &&= \gamma^{-1}\left(x\right) = x^2,\cr
h\left(x\right) &&= x^{-1},\cr
y\left(x\right) && = -  \sqrt{- \frac{2 \epsilon}{3}} = {\rm const.}
\end{eqnarray}
We evaluate the full equation of motion (\ref{eq:dimeom}) for a fiducial case of $\epsilon = - 0.435$, corresponding to the best-fit value of $w = - 1.29$ from the analysis of Di Valentino, {\it et al.} \cite{DiValentino:2016hlg}. Figure \ref{fig:attractor1} shows $y(x)$ vs $x$ for a variety of initial conditions, showing the attractor behavior of the solution (\ref{eq:attractor}). Figure \ref{fig:attractor2} shows the same attractor solution as a function of scale factor $a$ instead of as a function of the field, showing rapid convergence to the attractor solution. 

%%%%%%%%%%%%%%%%%%%%%%%%%%%%%%%%%%%%%%%%%%%%%%%%%%%%%%%%%%%%%%%%%%%%%%%%%%%%%%%%%%%%%%%%%%%%%%%%%
%%%%%%%%%%%%%%%%%%%%%%%%%%%%%%%%%%%%%%%%%%%%%%%%%%%%%%%%%%%%%%%%%%%%%%%%%%%%%%%%%%%%%%%%%%%%%%%%%
%%%%%%%%%%%%%%%%%%%%%%%%%%%%%%%%%%%%%%%%%%%%%%%%%%%%%%%%%%%%%%%%%%%%%%%%%%%%%%%%%%%%%%%%%%%%%%%%%
%%%%%%%%%%%%%%%%%%%%%%%%%%%%%%%%%%%%%%%%%%%%%%%%%%%%%%%%%%%%%%%%%%%%%%%%%%%%%%%%%%%%%%%%%%%%%%%%%

\subsection{Cosmological Boundary Conditions}
\label{sec:BoundaryConditions}

With attractor behavior established, we now consider the question of cosmological boundary condition. Our exact solution (\ref{eq:gensol1},\ref{eq:gensol2},\ref{eq:gensol3}) only applies to a {\it single-component} universe, {\it i.e.} it is a good approximation in the limit that Dark Energy dominates the cosmological energy density, and therefore the dynamics. However, in the presence of Dark Matter, the Dark Energy will be subdominant at high redshift, with the transition from matter- to phantom-domination happening at an approximate redshift of $z \sim 1$. We must therefore consider the dynamics of the field in the limit of matter-domination, which sets the boundary condition for the field evolution when the phantom energy dominates, at $z < 1$. 

\begin{figure}
\begin{center}
\includegraphics[width=0.8\textwidth]{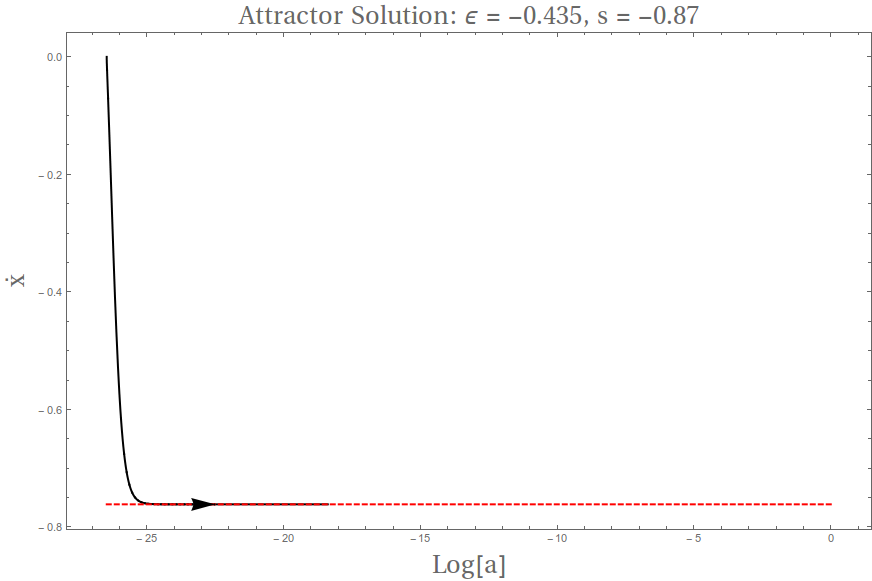}
\end{center}
\caption{Field evolution with a cosmological boundary condition $\dot\phi \rightarrow 0$. (Scale factor is in arbitrary units.) The dotted (red) line represents the late-time attractor solution. }
\label{fig:cosmobound}
\end{figure}

Consider the limit of large field, $x \gg 1$, so that the dimensionless warp factor (\ref{eq:dimf}) and potential (\ref{eq:dimv}) become
\begin{eqnarray}
&&g(x) \frac{3}{2 \epsilon} \left(1 - x^4\right) \rightarrow - \frac{3}{2 \epsilon} x^4,\cr
&&v(x) = x^{-2} \left(1 - \frac{2 \epsilon}{3} \frac{1}{1 + x^2}\right) \rightarrow x^{-2}.
\end{eqnarray}
We consider by {\it ansatz} a solution of the form $y\left(x\right) \rightarrow 0$. It is straightforward to identify subdominant terms in the equation of motion (\ref{eq:dimeom}),
\begin{eqnarray}
&&\frac{g'}{g} \rightarrow \frac{4}{x} \rightarrow 0,\cr
&&\frac{g'}{g^2} \rightarrow 0,\cr
&&y\left(x\right) y'\left(x\right) \rightarrow 0,\cr
&&c_S^2 = \gamma^{-2} = 1 + g\left(x\right) y\left(x\right) \rightarrow 1.
\end{eqnarray}
In this limit, the solution is a slowly rolling scalar,
\begin{equation}
\label{eq:SR}
3 H(t){\dot x} - v'\left(x\right) = 0,
\end{equation}
where $H\left(t\right)$ is determined by the scaling of the dominant energy component, either matter or radiation. Taking
\begin{equation}
H\left(t\right) = \frac{2}{3 \left(1 + w\right)} t^{-1},
\end{equation}
and $v\left(x\right) = x^{-2}$, the solution to (\ref{eq:SR}) is 
\begin{equation}
x\left(t\right) \propto \left(1 - C t^2\right)^{1/4} \rightarrow \mathrm{const.}
\end{equation}
This solution is confirmed by direct numerical integration of the full equation of motion (\ref{eq:dimeom}). We therefore have a boundary condition of a frozen field with $c_S = 1$ in the very early universe, which then relaxes to the attractor solution once the phantom component becomes dominant. Figure \ref{fig:cosmobound} shows field relaxation to the attractor solution with a cosmological boundary condition, $\dot\phi \rightarrow 0$, showing that relaxation occurs rapidly, in less than a Hubble time. 

%%%%%%%%%%%%%%%%%%%%%%%%%%%%%%%%%%%%%%%%%%%%%%%%%%%%%%%%%%%%%%%%%%%%%%%%%%%%%%%%%%%%%%%%%%%%%%%%%
%%%%%%%%%%%%%%%%%%%%%%%%%%%%%%%%%%%%%%%%%%%%%%%%%%%%%%%%%%%%%%%%%%%%%%%%%%%%%%%%%%%%%%%%%%%%%%%%%

\subsection{Vacuum Stability in Arbitrary Gauge}
\label{sec:ArbitraryGauge}

We have shown that, considered in the fluid rest frame, the energy density (\ref{eq:rho}) (and therefore the field Hamiltonian) is bounded from below, with a stable dynamical attractor solution corresponding to $p = w \rho$, which $w = {\mathrm const.} < -1$. The expression 
\begin{equation}
\rho = 2 X {\mathcal L}_X - {\mathcal L}
\end{equation}
is manifestly a coordinate scalar, and independent of gauge. However, the {\it Hamiltonian} is a coordinate-{\it dependent} object, corresponding to the time-evolution operator in a particlar foliation of the spacetime. For a perfect fluid, the Hamiltonian corresponds exactly to the energy density (\ref{eq:rho2}) in the rest frame of the fluid, where $T^{\mu}{}_{\nu} = {\mathrm diag}\left(\rho, -p, -p, -p\right)$, and
\begin{equation}
{\mathcal H} \equiv T^0{}_{0} = \rho.
\end{equation}
The fluid four-velocity in a general coordinate frame can be written as
\begin{equation}
u^\mu = \frac{\partial^\mu \phi}{\sqrt{2 X}},
\end{equation}
which is by construction timelike and unit normalized, $u^\mu u_\mu = +1$. (Note that timelike $u^\mu$ automatically implies that the kinetic term $X$ is positive-definite). The corresponding stress-energy is then
\begin{eqnarray}
T^{\mu \nu} &=& \left(\rho + p\right) u^\mu u^\nu - g^{\mu \nu} p\cr
&=& {\mathcal L}_X \partial^\mu \phi \partial^\nu \phi - g^{\mu \nu} {\mathcal L}\cr
&=& - \frac{\partial^\mu \phi \partial^\nu \phi}{\sqrt{1 + 2 f\left(\phi\right) X}} - g^{\mu \nu} {\mathcal L}.
\end{eqnarray}
We can then write the Hamiltonian in a general coordinate frame as
\begin{eqnarray}
{\mathcal H} &=& V\left(\phi\right) - \frac{1}{f\left(\phi\right)} + \frac{1 + f\left(\phi\right) \left(\partial^i \phi\right) \left(\partial_i \phi\right)}{f\left(\phi\right) \sqrt{1 + 2 f\left(\phi\right) X}}\cr
&=& \rho - \frac{\left(\nabla \phi\right)^2}{c_S}.\label{eq:generalH}
\end{eqnarray}
where $\left(\partial^i \phi\right) \left(\partial_i \phi\right) = - \left(\nabla \phi\right)^2$ denotes a sum over spatial indices. This reduces trivially to Eq. (\ref{eq:rho2}) in the limit of zero field gradient $\partial^i \phi = 0$, {\it i.e.} the rest frame of the fluid. This is in general not positive definite, since a negative Hamiltonian density can be found for a field configuration with gradient
\begin{equation}
(\nabla \phi)^2 > \rho c_s = \frac{1}{f\left(\phi\right)} \left(1 - c_S\right) + c_S V\left(\phi\right).
\label{eqn:negcondition}
\end{equation}
For the solution (\ref{eq:attractor}), this condition is especially simple, since
\begin{equation}
\frac{\rho}{3 M_{\rm P}^2} = h^2\left(x\right) = \gamma\left(x\right) = \frac{1}{c_S},
\end{equation}
so that 
\begin{equation}
c_S \rho = 3 M_P^2 H_0^2 = \mathrm{const.},
\end{equation} 
and ${\mathcal H} < 0$ for
\begin{equation}
(\nabla \phi)^2 > 3 M_P^2 H_0^2.
\end{equation}
While this relation shows that the Hamiltonian can be negative for sufficiently large field gradient, it does not show that the Hamiltonian is in fact {\it unbounded}. We show this below for a general NEC-violating perfect fluid, and apply the result to the specific model considered here.

The general case was shown by Sawiki and Vikman in Ref. \cite{Sawicki:2012pz}. Consider a perfect fluid with stress-energy violating the Null Energy Condition, such that there exists a null congruence $n^\mu$ such that
\begin{equation}
T_{\mu\nu} n^\mu n^\nu < 0.
\end{equation}
Take the fluid four-velocity to by given by a timelike congruence $u^\mu$. The rest-frame energy density $\rho$ is then given by
\begin{equation}
\rho = T_{\mu\nu} u^\mu u^\nu = 2 X {\mathcal L}_X - {\mathcal L}.
\end{equation}
This is a coordinate-invariant scalar, and is valid in any reference frame. However, it is only equal to the {\it Hamiltonian} in the rest frame of the fluid, $u^\mu = (1, 0, 0, 0)$. Now consider an arbitrary coordinate frame defined by a timelike congruence $v^\mu:\ v^\mu v_\mu = +1$. The Hamiltonian defined by this rest frame is given by
\begin{equation}
{\mathcal H} = T_{\mu\nu} v^\mu v^\nu.
\end{equation}
We now construct $v^\mu$ as a linear combination of the fluid four-velocity $u^\mu$ and null vector $n^\mu$,
\begin{equation}
\label{eq:defvmu}
v^\mu \equiv \alpha u^\mu + \beta n^\mu,
\end{equation}
where $\alpha$ and $\beta$ are constants, and $n^\mu$ is normalized such that $u^\mu n_\mu = +1$. Since $v_\mu$ is by definition unit normalized,
\begin{eqnarray}
v^\mu v_\mu = +1 &&= \alpha^2 u^\mu u_\mu + 2 \alpha \beta u^\mu n_\mu + \beta^2 n^\mu n_\mu\cr
&& = \alpha^2 + 2 \alpha \beta,
\end{eqnarray}
so that
\begin{equation}
\beta = \frac{1 - \alpha^2}{2 \alpha},
\end{equation}
and
\begin{equation}
v^\mu = \alpha u^\mu + \frac{1 - \alpha^2}{2 \alpha} n^\mu.
\end{equation}
The corresponding Hamiltonian is then
\begin{equation}
\label{eq:genHam}
{\mathcal H} = \alpha^2 T_{\mu\nu} u^\mu u^\nu + (1 - \alpha^2) T_{\mu\nu} u^\mu n^\nu + \frac{(1 - \alpha^2)^2}{4 \alpha^2} T_{\mu\nu} n^\mu n^\nu.
\end{equation}
Note that if $n^\mu$ satisfies the NEC, the last term is zero or positive-definite, but if $n^\mu$ violates the NEC, it is negative,
\begin{equation}
T_{\mu\nu} n^\mu n^\nu < 0.
\end{equation}
We are free to take the limit that $v^\mu$ is arbitrarily close to the light cone, $v^\mu \rightarrow n^\mu$, which corresponds to the limit $\alpha \rightarrow 0$. Then the Hamiltonian approaches
\begin{equation}
{\mathcal H} \rightarrow \frac{1}{4 \alpha^2} T_{\mu\nu} n^\mu n^\nu \rightarrow -\infty.
\end{equation}
The vector $v^\mu$ remains timelike and unit normalized, but the Hamiltonian is unbounded from below. For the particular case of the scalar field Lagrangian (\ref{eq:PhanLag}), 
\begin{eqnarray}
 T\indices{_\mu _\nu}u^\mu u^\nu &&= 2X \mathcal{L}_X - \mathcal{L} = \rho\\
T\indices{_\mu _\nu} u^\mu n^\nu &&= 2X\mathcal{L}_X \underbrace{(u_\alpha n^\alpha)}_1 \underbrace{(u_\alpha u^\alpha)}_1 - \mathcal{L} \underbrace{(u_\alpha n^\alpha)}_1 = 
2X\mathcal{L}_X - \mathcal{L} = \rho \\
T\indices{_\mu _\nu} n^\mu n^\nu &&= 2X\mathcal{L}_X
\end{eqnarray}
Then Eq. (\ref{eq:genHam}) reduces to 
\begin{equation}
{\mathcal H} = \rho + \frac{(1-\alpha^2)^2}{2 \alpha^2} X \mathcal{L}_X \rightarrow - \frac{1}{2 \alpha^2} \frac{X}{c_S},\ \alpha \rightarrow 0.
\end{equation}
Therefore, while the field Hamiltonian is bounded and well-behaved in the rest frame of the fluid, there exists a proper Lorentz frame for which the Hamiltonian appears {\it unbounded} from below.

\section{Conclusions}
\label{sec:Conclusions}

Current data suggest a tension between high-redshift constraints on the Hubble parameter $H_0$ from Cosmic Microwave Backround measurements \cite{Ade:2015xua}, and low-redshift constraints from Type-Ia supernovae \cite{Riess:2016jrr}. While still not compelling when viewed in terms of Bayesian evidence, this discrepancy suggests the need for inclusion of extended cosmological parameters beyond $\Lambda$CDM. Perhaps the conceptually simplest way to reconcile a low value of $H_0$ at high redshift with a high value at low redshift is Dark Energy which violates the Null Energy Condition, such that the expansion rate $H$ {\it increases} with expansion. Such ``Phantom'' Dark Energy (PDE) is characterized by Equation of State $p < - \rho$; while appealing from a parametric standpoint, PDE presents serious problems from a model-building standpoint. In particular, a phantom equation of state in a scalar field theory typically requires a wrong-sign kinetic term, which implies negative energy and a vacuum unstable to particle production \cite{Carroll:2003st,Cline:2003gs,Dubovsky:2005xd}. 
 
In this paper, we construct a Lagrangian with phantom equation of state $p < - \rho$, based on a Dirac-Born-Infeld (DBI) Lagrangian, with a wrong-sign kinetic term
\begin{equation}
{\mathcal L} = -\frac{1}{f\left(\phi\right)} \sqrt{1 + 2 f\left(\phi\right) X} + \frac{1}{f\left(\phi\right)} - V\left(\phi\right).
\end{equation}
For appropriate choices of $f\left(\phi\right)$ and $V\left(\phi\right)$, the comoving energy density $\rho$ is bounded from below,
\begin{equation}
\rho = \frac{1}{f\left(\phi\right) \sqrt{1 + 2 f\left(\phi\right) X}} - \frac{1}{f\left(\phi\right)} + V\left(\phi\right).
\end{equation}

We show by construction that it is possible to construct a Lagrangian with exact solution for homogeneous field modes such that $w \equiv p / \rho = {\mathrm const.} < -1$. These solutions correspond to a scalar field rolling {\it up} an approximately quadratic potential (Fig. \ref{fig:potential}). We show that these solutions correspond to a dynamical attractor in a cosmological background, and consider early-universe boundary conditions for the phantom field. We find field dynamics such that, in a matter-dominated phase at high redshift, the field is ``frozen'', and only becomes dynamical when the universe transitions to Dark Energy domination at low redshift. At that point, the field dynamics transitions to the attractor dynamics in less than a Hubble time, approaching constant equation of state $w < -1$. 

Since the DBI-type Lagrangian does not contain higher time-derivatives of the scalar field,  the theory automatically avoids any Ostrogradsky instabilities due non-local interactions \cite{Ostrogradsky:1850fid}.\footnote{See {\it e.g.} Ref. \cite{Woodard:2015zca} for a review of Ostrogradsky's theorem in the context of modern quantum field theories.} Despite these attractive features, the model is nonetheless pathological: While the field Hamiltonian is well-behaved in the {\it rest frame} of the fluid, it is dependent on the spacetime foliation. In particular, there in general exist proper Lorentz frames for which the Hamiltonian is unbounded from below \cite{Sawicki:2012pz}. This is distinct from the Unruh effect \cite{Unruh:1976db} in that it occurs for {\it inertial}, rather than accelerated observers. This is not necessarily an issue with repect to classical cosmological evolution, since there is no instability in the cosmological rest frame, and the classical gradient instability is nonlinear. However, quantum-mechanically, highly boosted momentum states will inevitably sample the region of phase space for which the field Hamiltonian can become arbitrarily negative \cite{Easson:2016klq}. Thus, while such models are attractive phenomenological descriptions of Phantom Dark Energy, they remain inconsistent as realizations of a fully fundamental theory. We note that the model considered here is purely phenomenological: wrong-sign Lagrangians are not, for example, typical of string-theory constructions \cite{Chatterjee:2015uya}. It is an interesting question whether or not Lagrangians of the type considered here can be embedded in a self-consistent UV-complete theory.

\section*{Acknowledgments}

GB acknowledges support from the MEC and FEDER (EC) Grants SEV-2014-0398, FIS2015-72245-EXP, and FPA2014-54459 and the Generalitat Valenciana under grant PROME-TEOII/2013/017. GB acknowledges partial support from the European Union FP7 ITN INVISIBLES MSCA PITN-GA-2011-289442 and InvisiblesPlus (RISE) H2020-MSCA-RISE-2015-690575. WHK thanks the University of Valencia and the Nordic Insitute for Theoretical Astrophysics in Stockholm for generous hospitality and support. WHK is supported by the National Science Foundation under grants NSF-PHY-1417317 and NSF-PHY-1719690. The authors thank Dragan Huterer and Richard Woodard for helpful conversations, and Alexander Vikman for comments on an earlier version of this paper.  

\bibliographystyle{JHEP}
\bibliography{paper}

\end{document}